\documentclass[twoside]{photon2007}
\usepackage[latin1]{inputenc}
\usepackage[dvips]{graphicx,epsfig,color}
\usepackage{wrapfig,rotating}
\usepackage{amssymb,amsmath,array}
\usepackage{bm}

\begin{document}
\title{
Virtual Photon Structure Functions to NNLO in QCD } 
\author{Takahiro Ueda$^1$~, Tsuneo Uematsu$^2$ and Ken Sasaki$^3$
\thanks{Presented by Ken Sasaki.}
\vspace{.3cm}\\
1-  High Energy Accelerator Research Organization (KEK)\\
Tsukuba 305-0801, Japan 
\vspace{.1cm}\\
2-  Dept. of Physics, Graduate school of Science, Kyoto University\\
Yoshida, Kyoto 606-8501, Japan
\vspace{.1cm}\\
3-  Dept. of Physics, Faculty of Engineering, Yokohama National University\\
Yokohama 240-8501, Japan
}


\maketitle

\begin{abstract}
\begin{picture}(0,0)
\put(330,220){KUNS-2121}
\put(330,210){YNU-HEPTh-07-103}
\end{picture}%
The unpolarized virtual photon structure
functions $F_2^\gamma(x,Q^2,P^2)$ and $F_L^\gamma(x,Q^2,P^2)$  are investigated in perturbative QCD
for the kinematical region $\Lambda^2 \!\ll \!P^2 \!\ll\! Q^2$, where $-Q^2(-P^2)$  is the mass squared 
of the probe (target) photon and $\Lambda$ is the QCD scale  parameter.
In the framework of operator product expansion supplemented by the renormalization group  method, 
the definite predictions are derived for the moments of $F_2^\gamma(x,Q^2,P^2)$
up to the next-to-next-to-leading order (the order $\alpha\alpha_s$) and for the 
moments of $F_L^\gamma(x,Q^2,P^2)$  up to the next-to-leading order (the order $\alpha\alpha_s$).
\end{abstract}

\section{Introduction}
\begin{figure}[h]
\centerline{\includegraphics[width=0.9\columnwidth]{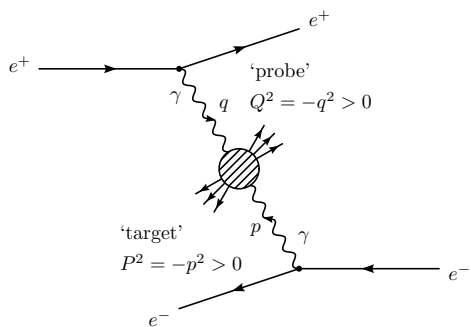}}
\caption{Deep-inelastic scattering on a virtual photon in the $e^+e^- $collider experiments}\label{Fig:two-photon}
\end{figure}
In $e^+e^-$ collision  experiments, the cross section for the two-photon 
processes $e^+e^-\rightarrow e^+e^- \!+\! {\rm hadrons}$,  shown in 
Fig.\ref{Fig:two-photon},   dominates  over other processes such as the 
annihilation process $e^+e^-\rightarrow \gamma^* \rightarrow {\rm hadrons}$
at high energies. 
In particular, the two-photon processes in the double-tag events, where 
one of the virtual photon is very far off shell (large $Q^2\!\equiv\! -q^2$) while the other 
is close to the mass shell (small $P^2\!\equiv\! -p^2$), 
can be viewed as  deep-inelastic electron-photon
scattering  and provide us the information on  the  structure of the photon.
 The unpolarized (spin-averaged) photon structure functions $F_2^\gamma(x,Q^2)$ and 
 $F_L^\gamma(x,Q^2)$ of the real photon ($P^2\approx 0$) have been studied in perturbative QCD (pQCD).
A pioneering work was done by Witten \cite{Witten} in which he derived the leading order (LO) QCD contributions to $F_2^\gamma$ and $F_L^\gamma$. A few years later, the next-to-leading order (NLO)  corrections to $F_2^\gamma$ were calculated \cite{BB}.

A unique and interesting feature  of the photon structure functions is
that, in contrast with the nucleon case, the target mass squared $-P^2$ 
is not fixed but can take various values and that the structure functions show 
different behaviors depending on the values of $P^2$. The photon has two characters: 
The photon couples directly to quarks (pointlike nature) and sometimes 
it behaves as vector bosons (hadronic nature). Thus the structure function 
$F_2^\gamma(x,Q^2)$ of real photon ($P^2\!=\!0$) is composed of a 
pointlike piece and a hadronic piece. 
The pointlike part, can be calculated, in principle, in a perturbative method. 
On the other hand, the hadronic part, can only be computed by some nonperturbative method
like  lattice QCD, or estimated by vector meson dominance model. 
The LO  contribution to $F_2^\gamma(x,Q^2)$, which behaves as 
$1/\alpha_s(Q^2)\sim\ln (Q^2/\Lambda^2)$,  
comes from the pointlike part, while the NLO corrections result from 
both the pointlike and hadronic parts. In terms of the moments,  the hadronic 
energy-momentum tensor operator comes into play at n \!=\! 2. Because of 
the conservation of this operator,  the hadronic part gives a finite but 
perturbatively incalculable contribution at $n\!=\!2$.
The fact that definite information on the NLO second moment is missing prevents us from fully predicting the shape and magnitude of the structure function of $F_2^\gamma(x,Q^2)$ up to 
the order ${\cal O}(\alpha)$

The situation changes significantly when we analyze the structure function of a virtual photon 
with $P^2$ much larger than the QCD parameter $\Lambda^2$\cite{UW2}. More specifically, we consider the following kinematical region, 
\begin{equation}
\Lambda^2 \ll P^2 \ll Q^2 \label{Kinematical}~.
\end{equation}
In this region, 
the hadronic component of the photon can also be dealt with {\it perturbatively} 
and thus a definite prediction of the whole structure function, its shape and magnitude,  
may become possible.  In fact,   
the virtual photon structure function $F_2^\gamma(x,Q^2,P^2)$ in the kinematical region
(\ref{Kinematical}) was calculated  in the LO (the order $\alpha/\alpha_s$) \cite{UW1} and in the NLO  (the order $\alpha$)~\cite{UW2,Rossi}, and the longitudinal  
structure function $F_L^\gamma(x,Q^2,P^2)$ in the LO (the  order $\alpha$)~\cite{UW2} 
without any unknown parameters.

In this talk  I report our  investigation  \cite{USU} of the virtual photon structure functions $F_2^\gamma(x,Q^2,P^2)$ up to the next-to-next-to-leading order
(NNLO)  (the order $\alpha\alpha_s$) and $F_L^\gamma(x,Q^2,P^2)$ up to the NLO (the order $\alpha\alpha_s$) in the kinematical region (\ref{Kinematical}).

\section{$F_2^\gamma(x,Q^2,P^2)$ up to NNLO}
We have used the framework of the operator product 
expansion (OPE) supplemented by the renormalization group (RG) method. We find that
the $n$-th moment of  $F_2^\gamma(x,Q^2,P^2)$ for the kinematical region 
(\ref{Kinematical}) is expressed, up to NNLO,  as
\begin{eqnarray}
&&\hspace{-0.8cm}\int_0^1 dx x^{n-2}F_2^\gamma(x,Q^2,P^2)  
/\Bigl(  \frac{\alpha}{4\pi}\frac{1}{2\beta_0} \Bigr)\nonumber\\
&=&\hspace{-0.5cm}
\Biggl\{\frac{4\pi}{\alpha_s(Q^2)}\sum_{i}{\cal L}^n_i
\biggl[1-\left(\frac{\alpha_s(Q^2)}{\alpha_s(P^2)}\right)^{d_i^n+1}
\biggr]\nonumber\\
&&\quad+\sum_{i}{\cal
A}_i^n\biggl[ 1-\left(\frac{\alpha_s(Q^2)}{\alpha_s(P^2)}\right)^{d_i^n}\biggr]\nonumber\\
&& \quad+\sum_{i}{\cal
B}_i^n\biggl[ 1-\left(\frac{\alpha_s(Q^2)}{\alpha_s(P^2)}\right)^{d_i^n+1}\biggr]
+{\cal C}^n \nonumber\\
&&\hspace{-0.3cm}+\frac{\alpha_s(Q^2)}{4\pi}\biggl(\sum_{i}{\cal
D}_i^n\biggl[ 1-\left(\frac{\alpha_s(Q^2)}{\alpha_s(P^2)}\right)^{d_i^n-1}\biggr]\nonumber\\
&&+\sum_{i}{\cal
E}_i^n\biggl[ 1-\left(\frac{\alpha_s(Q^2)}{\alpha_s(P^2)}\right)^{d_i^n}\biggr]
\nonumber\\
&& +\sum_{i}{\cal
F}_i^n\biggl[ 1-\left(\frac{\alpha_s(Q^2)}{\alpha_s(P^2)}\right)^{d_i^n+1}\biggr]
+{\cal G}^n \biggr) \nonumber\\
&&\hspace{-0.3cm}+{\cal O}(\alpha_s^2)
 \Biggr\},\quad {\rm with}\quad i=+, -, NS~,
\label{master1}
\end{eqnarray}
where $d_i^n=\lambda_i^n/2\beta_0$ and $\lambda_i^n (i=+, -, NS)$ denotes the eigenvalues of 1-loop anomalous dimension matrices. 
The terms with ${\cal L}^n_i$ are the LO ($\alpha/\alpha_s$) contributions \cite{Witten}. The 
NLO ($\alpha$) corrections are the terms with ${\cal A}^n_i$, ${\cal B}^n_i$ and ${\cal C}^n$ 
\cite{BB,UW2}. The coefficients  ${\cal D}^n_i$, ${\cal E}^n_i$,  ${\cal F}^n_i$ and  ${\cal G}^n$  
give the NNLO ($\alpha\alpha_s$) corrections and they are new. 
The explicit expressions of  ${\cal D}^n_i$, ${\cal E}^n_i$,  ${\cal F}^n_i$ and  ${\cal G}^n$ 
are given in Eqs.(2.34)-(2.37) of Ref.\cite{USU} and they are written in terms of
the 1-, 2- and 3-loop anomalous dimensions, the 1- and 2-loop 
coefficient functions and the 1- and 2-loop photon matrix elements of hadronic operators.  

For the 3-loop anomalous dimensions, we could use
the recently calculated results of the three-loop anomalous dimensions for the quark and gluon 
operators~\cite{MVV1,MVV2} and  
of the three-loop photon-quark and photon-gluon splitting functions \cite{MVV3}. 
The 2-loop photon matrix elements of hadronic operators were derived from  the results of  the two-loop operator matrix elements calculated up to the finite terms \cite{MSvN}
by changing color-group factors.
 
We examine the sum rule of  $F_2^\gamma(x,Q^2,P^2)$, i.e., the second moment,  
numerically. 
The NNLO corrections are found to be  
$7\% \sim10\%$ of the sum of the LO and NLO contributions, when
 $P^2\!=\!1 {\rm GeV}^2$ and  $Q^2\!=\!30\sim 100{\rm GeV}^2$ or 
$P^2\!=\!3 {\rm GeV}^2$ and  $Q^2\!=\!100{\rm GeV}^2$,  and $n_f$ is  three or four. 

Next we perform the inverse Mellin transform of  (\ref{master1}) to obtain $F_2^\gamma$ as a function of $x$ .
 The $n$-th moment is denoted as
\begin{equation}
 M_2^\gamma(n,Q^2,P^2)=\int_0^1 dx~x^{n-1}\frac{F_2^\gamma(x,Q^2,P^2)}{x}~.
 \label{mom}
 \end{equation}
 Then by inverting the moments (\ref{mom}) we get
 \begin{eqnarray}
 &&\hspace{-1cm}\frac{F_2^\gamma(x,Q^2,P^2)}{x}\nonumber \\
&&\hspace{-1cm}\quad  =\frac{1}{2\pi i}
 \int_{C-i\infty}^{C+i\infty}dn~ x^{-n} M_2^\gamma(n,Q^2,P^2)~,
 \end{eqnarray}
where the integration contour runs to the right of all singularities
of $ M_2^\gamma(n,Q^2,P^2)$ in the complex $n$-plane. 

\begin{figure}[h]
\centerline{\includegraphics[width=1.0\columnwidth]{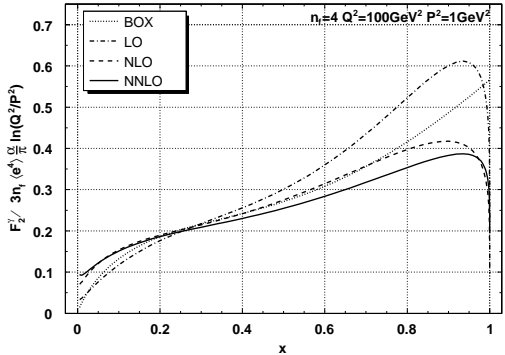}}
\caption{Virtual photon structure function $F_2^\gamma(x,Q^2,P^2)$ for
$Q^2\!=\!100$ GeV$^2$ and $P^2\!=\!1$ GeV$^2$ with 
 $n_f\!=\!4$ and $\Lambda\!=\!0.2$ GeV.}\label{F2_4_100_1_200}
\end{figure}

The LO, NLO and NNLO QCD results, as well as the box contribution,  for the case of 
$Q^2\!=\!100$ GeV$^2$ and $P^2\!=\!1$ GeV$^2$ with 
 $n_f\!=\!4$,  are shown in Fig.\ref{F2_4_100_1_200}. We observe that there exist 
 notable NNLO QCD corrections at larger $x$. The corrections are negative and the NNLO 
 curve comes below the NLO one in the region $0.3\!\lesssim\!x\!<\!1$. At the lower $x$ region, 
 $0.05\!\lesssim\!x\!\lesssim\!0.3$, the NNLO corrections to the NLO results are found to be negligibly small.

\section{$F_L^\gamma(x,Q^2,P^2)$ up to NLO}
The formula for the $n$-th moment of the longitudinal structure function 
$F_L^\gamma(x,Q^2,P^2)$ can be  obtained from (\ref{master1}) only by replacing the hadronic and photonic coefficient functions $\bm{C}_{2,n}(1,\alpha_s)$ and 
$C_{2,n}^\gamma (1,\alpha_s,\alpha)$ with the longitudinal 
counterparts $\bm{C}_{L,n}(1,\alpha_s)$ and 
$C_{L,n}^\gamma (1,\alpha_s,\alpha)$, respectively.
Since there is  no contribution of the tree diagrams to the 
hadronic longitudinal coefficient functions (and thus we get $\bm{C}^{(0)}_{L,n}=\bm{0}$ 
in the expansion of $\bm{C}_{L,n}(1,\alpha_s)$), the moments of $F_L^\gamma$ starts 
at the order $\alpha$. The $n$-th moment is given as follows:
\begin{eqnarray}
&&\hspace{-0.6cm}\int_0^1 dx x^{n-2}F_L^\gamma(x,Q^2,P^2)
/\Bigl(  \frac{\alpha}{4\pi}\frac{1}{2\beta_0} \Bigr)  \nonumber\\
&&\hspace{-0.6cm}=\Biggl\{\sum_{i}{\cal
B}_{(L),i}^n\biggl[1-\left(\frac{\alpha_s(Q^2)}{\alpha_s(P^2)}\right)^{d_i^n+1}\biggr]
+{\cal C}_{(L)}^n \nonumber\\
&&\hspace{-0.5cm}+\frac{\alpha_s(Q^2)}{4\pi}\biggl(\sum_{i}{\cal
E}_{(L),i}^n\biggl[1-\left(\frac{\alpha_s(Q^2)}{\alpha_s(P^2)}\right)^{d_i^n}\biggr]
\nonumber\\
&&\hspace{-0.5cm}+\sum_{i}{\cal
F}_{(L),i}^n\biggl[1-\left(\frac{\alpha_s(Q^2)}{\alpha_s(P^2)}\right)^{d_i^n+1}\biggr]
+{\cal G}_{(L)}^n \biggr) \nonumber\\
&&\hspace{-0.5cm}+{\cal O}(\alpha_s^2) 
 \Biggr\}~,\quad {\rm with}\quad i=+, -, NS~,
\label{masterL}
\end{eqnarray}
The coefficients ${\cal B}^n_{(L),i}$ and ${\cal C}^n_{(L)}$ 
represent the LO terms~\cite{Witten,BB,UW2}, while the terms with 
${\cal E}^n_{(L),i}$, ${\cal F}^n_{(L),i}$ and ${\cal G}^n_{(L)}$ are the NLO 
($\alpha\alpha_s$) corrections and they are new. 
The explicit expressions of  ${\cal E}^n_{(L),i}$, ${\cal F}^n_{(L),i}$ and ${\cal G}^n_{(L)}$
are given in Eqs.(6.6)-(6.8) of Ref.\cite{USU}.

\begin{figure}[h]
\centerline{\includegraphics[width=1.0\columnwidth]{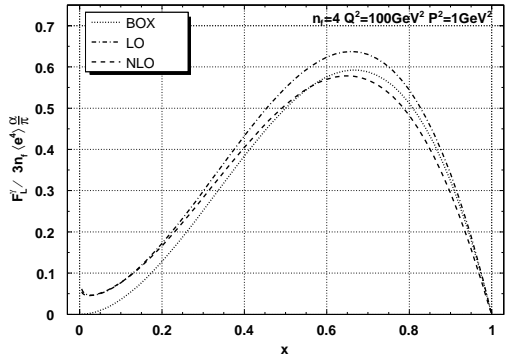}}
\caption{Longitudinal photon structure function $F_L^\gamma(x,Q^2,P^2)$ for
$Q^2\!=\!100$ GeV$^2$ and $P^2\!=\!1$ GeV$^2$ with 
 $n_f\!=\!4$ and $\Lambda\!=\!0.2$ GeV.}\label{FL_4_100_1_200}
\end{figure}

Inverting the moments (\ref{masterL}), we plot in Fig.\ref{FL_4_100_1_200} 
the longitudinal virtual photon structure function  $F_L^\gamma(x,Q^2,P^2)$
predicted by pQCD for the case of $n_f\!=\!4$, $Q^2\!=\!100$ GeV$^2$ and $P^2\!=\!1$ GeV$^2$.  We show three curves; the LO and NLO QCD results and
the Box (tree) diagram contribution. We see that the NLO QCD corrections are negative and the NLO curve comes below
the LO one  in the region  $0.2\!\lesssim\! x<1$.

\section{Conclusions}

We have investigated the unpolarized virtual photon structure
functions $F_2^\gamma(x,Q^2,P^2)$ and $F_L^\gamma(x,Q^2,P^2)$ 
for the kinematical region $\Lambda^2 \ll P^2 \ll Q^2    $ in QCD.
In the framework of the OPE supplemented by the RG method, 
we gave the definite predictions for the moments of $F_2^\gamma(x,Q^2,P^2)$
up to the NNLO (the order $\alpha\alpha_s$) and for the 
moments of $F_L^\gamma(x,Q^2,P^2)$  up to the NLO  (the order $\alpha\alpha_s$).
In the course of  our evaluation,
we utilized
the recently calculated results of
 the three-loop anomalous dimensions for the quark and gluon 
operators. Also we derived the photon matrix elements of hadronic operators  
up to the two-loop level.

The inverse Mellin transform of the moments was performed to 
express the structure functions $F_2^\gamma(x,Q^2,P^2)$ and $F_L^\gamma(x,Q^2,P^2)$
as  functions of $x$.
We found that there exist sizable NNLO  contributions for $F_2^\gamma$ at larger $x$.
The corrections are negative and the NNLO curve comes below the NLO one 
in the region  $0.3\lesssim x<1$.
 At lower $x$ region, $0.05\lesssim x \lesssim 0.3$, the NNLO corrections to the NLO
results are found to  be negligibly small. Concerning  $F_L^\gamma$, 
the NLO corrections  reduce the magnitude in the region 
 $0.2\lesssim x<1$.
 
\section*{Acknowledgments}
This research is supported in part by Grant-in-Aid
for Scientific Research  from the Ministry of Education, Culture, Sports, Science and Technology,
Japan No.18540267.


\begin{footnotesize}




\begin{thebibliography}{99}

\bibitem{Witten}
     E.~Witten, {Nucl.~Phys.} {\bf B120} 189 (1977).
     
 \bibitem{BB}
     W.A.~Bardeen and A.J.~Buras, {Phys. Rev.} {\bf D20} 166 (1979);   
    {Phys. Rev.} {\bf D21}  2041 (1980), Erratum.
    
\bibitem{UW2}
      T.~Uematsu and T.~F.~Walsh, {Nucl.~Phys.} {\bf B199} 93 (1982).
      

\bibitem{UW1}
      T.~Uematsu and T.~F.~Walsh, {Phys.~Lett.} {\bf B199} 263 (1981).
      
\bibitem{Rossi}
     G.~Rossi, {Phys. Rev.} {\bf D29}  852 (1984).

     \bibitem{USU}
      T.~Ueda, K.~Sasaki and T.~Uematsu, {Phys. Rev.} {\bf D75} 114009 (2007).
      
      
 
\bibitem{MVV1}
    S.~Moch, J.A.M.~Vermaseren and A.~Vogt, { Nucl.~Phys.} {\bf B688}, 101 (2004).


\bibitem{MVV2}
   A.~Vogt,  S.~Moch and J.A.M.~Vermaseren,  {Nucl.~Phys.} {\bf B691}, 129 (2004).

 
\bibitem{MVV3}
   A.~Vogt,  S.~Moch and J.A.M.~Vermaseren,  { Acta Phys.~Pol.} {\bf B37}, 683 (2006); 
arXiv:hep-ph/0511112.    
      

\bibitem{MSvN}
    Y.~Matiounine, J.~Smith and  W.~L.~van~Neerven, { Phys. Rev.} {\bf D57}, 6701 (1998).
      

\end{thebibliography}
%

\end{footnotesize}


\end{document}